\documentclass[11pt,oneside,letterpaper]{article}
\usepackage{amssymb}
\usepackage{amsmath,stackengine}
\usepackage[dvips]{graphicx}
\usepackage{setspace}
\usepackage{amsfonts}
\usepackage{fancyhdr}
\usepackage[table]{xcolor}
\usepackage{xcolor}
\usepackage{mathrsfs}
\usepackage{diagbox}
\usepackage{multirow}
\usepackage{graphicx}
\usepackage{rotating}
\usepackage{comment}
\usepackage{fancyvrb}
\usepackage{color}
\usepackage{empheq}
\usepackage{cite}
\usepackage{braket}
\usepackage{caption}
\usepackage[utf8]{inputenc}
\usepackage[c]{esvect}
\usepackage{mathabx}
\usepackage{framed}
\usepackage{scalerel}
\usepackage{tikz}
\usetikzlibrary{3d,calc}
\usetikzlibrary{patterns}
\usepackage{moresize}
\usetikzlibrary{decorations.pathmorphing, arrows.meta, positioning}
\usepackage{tcolorbox}
\usepackage[stable]{footmisc}
\usepackage[hidelinks,colorlinks=true,linkcolor=blue,citecolor=blue]{hyperref}
\usepackage{textcomp}
\usepackage[T1]{fontenc}  % Needed for \l and \L
\usepackage[utf8]{inputenc}  % Match your file encoding
\usepackage[polish, english]{babel} 

\tcbuselibrary{breakable}

\usepackage{mwe}

\definecolor{lightgray}{gray}{0.85}

\definecolor{darkgreen}{rgb}{0,0.5,0}
\definecolor{darkblue}{rgb}{0,0,0.6}
\definecolor{purple}{rgb}{0.4,.2,0.7}

\newcommand{\be}{\begin{equation}}
	\newcommand{\ee}{\end{equation}}

\newcommand{\ti}{{\tt{ t}}}

\usepackage{pdfsync}
\makeatletter
\newcommand*{\mybox}{\mathrel{%
			\raisebox{0.25ex}{$\Box$}}%
	} 
\makeatother
\DeclareMathOperator{\diag}{diag}

\def\be{\begin{eqnarray}}
	\def\ee{\end{eqnarray}}

\newcommand{\bea}{\begin{eqnarray}}
	\newcommand{\eea}{\end{eqnarray}}
\def\ben{\begin{equation}}
	\def\een{\end{equation}}

     \let\r=v

\def\be{\begin{equation}}
	\def\ee{\end{equation}}
\def\ba{\begin{array}}
	\def\ea{\end{array}}

\def\ba#1\ea{\begin{align}#1\end{align}}
\def\bs#1\es{\begin{split}#1\end{split}}

\allowdisplaybreaks  % allow page breaks in displayed eqs

\numberwithin{equation}{section}
\def \be {\begin{equation}}
	\def \ee {\end{equation}}
	
\def \JM#1 {{\color{blue}  JM: #1 }}
\def \AAl#1 {{\color{red}  AA: #1 }}

\DeclareRobustCommand{\rchi}{{\mathpalette\irchi\relax}}
\newcommand{\irchi}[2]{\raisebox{0.3ex}{$#1\chi$}} % inner command, used by \rchi

\begin{document}
	\onehalfspacing
	
	\begin{center}
		
		~
		\vskip5mm

					{\LARGE  {Time as a test-field:  the  no-boundary universe \\[2.5mm]  in motion and a smooth radiation  bounce 
	}}		
			\vskip10mm

			 Federico Piazza and  Siméon Vareilles
	
		{\it Aix Marseille Univ, Universit\'{e} de Toulon, CNRS, CPT, Marseille, France}
			%} 
		
		\vskip5mm

		%{\tt piazza@cpt.univ-mrs.fr }
		
		\href{mailto:piazza@cpt.univ-mrs.fr}{\tt piazza@cpt.univ-mrs.fr}, \ \ \ \ \href{mailto:vareilles@cpt.univ-mrs.fr}{\tt vareilles@cpt.univ-mrs.fr}
	\end{center}
	
	\vspace{4mm}
	
	\begin{abstract}
	The proper time of an observer can be introduced as a degree of freedom in quantum cosmology, additional to the existing fields. We review two arguments for using the Schr\"odinger equation to evolve the corresponding wavefunction. We restrict to solutions in which time acts as a component with negligible backreaction on the metric---that is, it plays the role of a test field.  
We apply this idea to various minisuperspace models. In the semiclassical regime we recover expected results:  the wavefunction peaks on the classical solution and, in models with a scalar field, the variance of $\zeta$ (a mini-superspace analogue of the comoving curvature perturbation)  is conserved. Applied to the no-boundary wavefunction, our model recovers the bouncing behavior of classical global de Sitter space, with small  corrections associated to the evolving variance of the wavefunction. Other bouncing solutions do not have any classical analogue. This is the case of a radiation dominated universe, which classically leads to a big-bang singularity but corresponds quantum mechanically to an $s$-wave scattering off a central potential of the form $-r^{-2/3}$. As much as the hydrogen atom, this potential is famously made stable by the Heisenberg uncertainty principle. We study the unitary evolution of the wavepacket numerically. 
%The \emph{radiation bounce}, as opposed to the \emph{no-boundary bounce}, does not have a classical counterpart as the corresponding classical solutions are singular.  
During the bounce, the uncertainty and the expectation value of the scale factor become comparable.  By selecting a large initial variance, the bounce can be made arbitrarily smooth, the mean value of the Hubble parameter correspondingly soft.

 	 \end{abstract}
%\vspace{.2in}
%\vspace{.3in}

\pagebreak
\pagestyle{plain}

\setcounter{tocdepth}{2}
{}
\vfill

\ \vspace{-2cm}
\newcommand{\deq}{{\overrightarrow {\Delta x}}^{\, 2}}
\renewcommand{\baselinestretch}{1}\small

%\newpage
\tableofcontents
%\vspace{0.05cm}
\begin{table}[h!]
    \centering
    \small
    \captionsetup{width=12cm}
    \caption*{Summary of notation and conventions. We use natural units, $c = \hbar = 1$.}
    \setlength{\tabcolsep}{2pt}
    \label{tab:notation}
    \renewcommand{\arraystretch}{1.2} 
    \begin{tabular}{l l @{\hspace{-1.5em}} c}
        \hline\hline
        \textbf{Symbol} & \textbf{Description} & \textbf{Dim. $[M]$} \\ 
        \hline
        $\ti$ & Proper time of the observer / clock variable & $-1$ \\
        $p_\ti$ & Momentum conjugate to proper time variable  & $+1$ \\
        $t, \lambda$ &  Global coordinate time, wordline parameter& $0$ \\
        $N, N^i$ & Lapse and shift functions & 0 \\
        $\sqrt{-g_{\lambda\lambda}}$ & line element per unit $\lambda$ $\left(\frac{d\ti}{d\lambda}= \sqrt{-g_{\lambda\lambda}}=\sqrt{g_{\mu \nu} \frac{d x^\mu}{d \lambda}\frac{d x^\nu}{d \lambda}}\right)$ & 0 \\
        $q^a$ & Gravitational and matter fields degrees of freedom (minisuperspace variables) & $0$ \\
        $\mathcal{H}(q^a)$ & Gravitational and matter fields Hamiltonian  & $0$ \\
        $\Psi_{WdW}(q^a)$ & Wheeler DeWitt wavefunction, $\mathcal{H} \Psi_{WdW} = 0$  &0 \\
        $\Psi(q^a; \ti)$ & Time-dependent wavefunction of the universe, satisfying the Schr\"odinger eq. & $0$ \\
        $g_{ab}, \mybox$ & Metric in minisuperspace and its covariant d'Alembertian operator & $0$ \\
        $H_\star$ & Characteristic Hubble parameter & $+1$ \\
        $\alpha$ & Dimensionless gravitational coupling ($\alpha = G_N H_\star^2$) & $0$ \\
        $S(q^a)$ & Hamilton-Jacobi function, which we take dependent only on $q^a$ and not on $\tt{t}$ & $0$ \\
        $\chi(q^a), \psi(q^a; \ti)$  & Time-independent and time dependent factors in the factorized  WKB ansatz &0\\
        $a, x $ & Scale factor and canonically normalized scale factor ($x = a^{3/2}$)  & $0$ \\
        $\phi$ & Scalar field  & $0$ \\
        $\zeta$ & Minisuperspace comoving curvature perturbation& $0$ \\
        $\epsilon$ & Slow-roll parameter & $0$ \\
        
        \hline\hline
    \end{tabular}
\end{table}

\renewcommand{\baselinestretch}{1.15}\normalsize

%\section{What state is the state of the universe in?}

\section{The Schr\"odinger equation in quantum cosmology}

The Wheeler-DeWitt (WdW) equation~\cite{DeWitt:1967yk} is famously timeless. To recover some familiar dynamics, one of the fields can effectively serve as  ``time"~\cite{DeWitt:1967yk}.  One should then interpret the WdW wavefunction in terms of  conditional probability---for the remaining fields to be in a certain configuration given that the \emph{time-field} takes on a specific value. It is worth asking whether such a field could be chosen to  behave like proper time---the thing  measured by any properly functioning clock~\cite{teitelboim1983proper}. In the classical theory, this is essentially a global issue. Proper time can be associated to any given worldline but it might fail to be defined globally if e.g. the worldlines intersect. 
In the following we discuss a proposal for introducing in the WdW equation a dynamical field that behaves like proper time. While the global issues  could be exacerbated by the quantum treatment, no obstruction appears to be present in the mini-superspace approximation. In this limit, we shall argue that proper time can be introduced at the full quantum mechanical level and suggest---perhaps with little surprise---that the corresponding time-dependent wavefunction of the universe satisfies a Schr\"odinger equation, 
\begin{equation} \label{intro_main}
i \partial_\ti \Psi(q^a, \ti) \, = \, {\cal H}\, \Psi(q^a, \ti)\, ,
\end{equation}
where $q^a$ are mini-superspace variables such as the scale factor and the matter fields, while ${\cal H}$ is the Hamiltonian---the same that evaluates zero on the WdW wavefunction \emph{without} time, ${\cal H} \Psi_{WdW}(q^a)=0$. Clearly,  $\Psi(q^a, \ti)$ and $\Psi_{WdW}(q^a)$ are different, albeit related, objects.

In what follows we present two main arguments for~\eqref{intro_main}, one based on the path integral formalism (Sec.~\ref{1.1}) and one based on a clock model Hamiltonian  (Sec.~\ref{1.2}). An independent, perhaps expected, endorsement of~\eqref{intro_main} comes from the behavior of the wavefunction $\Psi(q^a, \ti)$ in the semiclassical limit (e.g. away from turning points and potential singularities), which we show to be consistent  with the classical evolution in  Sec.~\ref{sec_semiclassical}. The probability associated to $\Psi(q^a, \ti)$ is the standard ``modulus squared" one of the Schr\"odinger equation. Without the known subtleties related to the interpretations of the  standard Wheeler DeWitt wavefunction $\Psi_{WdW}(q^a)$, here the probability density is conserved and positive definite (see  Sec.~\ref{sec_probability}). 

%Although for some the problem of time remains an elusive and almost mystical subject confined to philosophical debate, others have adopted a more pragmatic standpoint.
\vspace{-0.25cm}
\subsubsection*{Comparison with existing approaches}
The idea of using proper time and the Schr\"odinger equation in quantum gravity  is not new  (see e.g.~\cite{teitelboim1983proper,unruh1989time,Alexander:2012tq,Peter_2018,Malkiewicz:2022szx,Burns:2022fzs,Kaya:2022mnb,Bergeron:2025eda}). Of particular relevance to the present work are Refs.~\cite{Vachaspati:2006ki,Greenwood:2008ht,Wang:2009ay,Saini:2014qpa} where the Schr\"odinger equation has been employed to investigate and resolve black hole type singularities. In deparametrized formulation of gravity~\cite{Martin:2021dbz, Malkiewicz:2022szx, Mazde:2025zne,Vitenti:2026hgs} typically, one adopts a relational approach where the evolution of geometric variables is described relative to a reference field clock whose conjugate momentum appears linearly in the hamiltonian constraint. Hence, the resulting Hamiltonian generates evolution with respect to the chosen internal time. A common well-known example of this idea is given by Brown--Kucha\v{r} dust~\cite{Brown:1994py}, where the dust proper time plays the role of the clock and the Hamiltonian constraint becomes linear in the dust momentum. In this work, we adopt a closely related but minimal perspective. Rather than introducing an explicit dust reference field at the level of the action, we work directly in a framework where the physical time parameter is identified with proper time, corresponding to the choice of lapse $N=1$. As we will explicitely develop in Sections \ref{1.1} and \ref{1.2}, rather than introducing a spacetime-filling dust field at the superspace level, we attach a proper-time degree of freedom to an observer's worldline, described by a canonical pair $(t,p_t)$ governed by a reparametrization-invariant worldline action. This clock variable behaves classically as the proper time along the worldline and leads, in minisuperspace, to a Hamiltonian contribution linear in $p_t$. While the resulting quantum dynamics coincides with that obtained using standard Brown--Kucha\v{r} dust in minisuperspace (in the sense that the conjugate momentum of both time variables is the energy of the proper-time clock sector), the underlying interpretation differs in that time is associated with a localized observer clock rather than a spacetime-filling dust field. Other types of time have also been considered, among which unimodular time in a number of papers~\cite{Gielen_2020,Gielen_2022,Gielen_2023,Gielen_2025,Sahota:2025tih, Ried:2025ipl}. In this approach, the cosmological constant is promoted to a dynamical variable, canonically conjugate to a global time parameter related to the spacetime four-volume ($N=a^{-3}$). From this perspective, unimodular time provides an alternative choice of physical clock, distinct from matter-based clocks such as Brown--Kucha\v{r} dust or the worldline clock modelled here. The existence of several inequivalent choices of time, including matter clocks, worldline clocks, and unimodular time, makes manifest the multiple-choice problem of time in quantum gravity. Physical results such as the resolution of the singularity, indeed, appear to depend on the actual choice of time~\cite{Gielen_2020}. In this respect, however, the proper time of an observer, as emerging from the periodic microphysical phenomena along its worldline, seems the appropriate diagnostic for singularities, which are defined in classical gravity by geodesic incompleteness~\cite{Hawking:1973uf}, i.e. by the impossibility of extending a geodesic \emph{past some value of their proper time} (see also~\cite{Piazza:2025uxm}).   It is worth emphasizing the distinct roles played by the clock degree of freedom and the matter content driving the cosmological dynamics in the present work. In our construction, the proper-time clock is introduced as an additional canonical degree of freedom \emph{whose energy density is negligible compared to that of the dominant sector}. Its purpose is to provide a physical notion of time with respect to which quantum evolution can be defined without influencing the cosmological evolution itself.  As specific examples in this paper, we study models where the dominant component is a cosmological constant + spatial curvature (giving a quantum version of De Sitter space, see Sec.~\ref{sec_nb}) and a radiation fluid (Sec.~\ref{bounce}). To our knowledge, the use of a  dynamical subdominant time separated from the dominant component has never been made before.
\vspace{0.2cm}
\subsection{The  path integral point of view\footnote{See also~\cite{Lehners:2023yrj}, Secs. 2.2 and 2.3}} \label{1.1}

In the path integral formalism one can evaluate the transition amplitude between two spatial metrics $h_{ij}(\vec x)$ and $h'_{ij}(\vec x)$ by integrating over all four-geometries that have $h_{ij}$ and $h'_{ij}$ as spatial boundaries,
\begin{equation}\label{intro_prima}
\langle h'_{ij} |  h_{ij}\rangle = \int \mathcal{D}g \, e^{i I_{\rm EH}[g]}\, .
\end{equation}
As a function of $h'_{ij}$, this expression satisfies the WdW equation~\cite{Hartle:1983ai,Halliwell:1990qr}. There is no reference to time in the above expression. However, 
the four-geometries that we are integrating over ``know" about time. 
One way to make this explicit is to write them in the ADM form 
\begin{equation}
\label{ADM_metric}
ds^2 = -N^2 dt^2 + \gamma_{ij}\,(dx^i + N^i dt)(dx^j + N^j dt)\,.
\end{equation}
and fix the gauge $N=1$, $N^i=0$. In this gauge, the worldlines with constant $\vec x$ are geodesics and the coordinate time $\ti$ coincides with the proper time along them. The only degree of freedom left in this gauge is the spatial metric $\gamma_{ij}(t, \vec x)$ of the hypersurfaces of constant $\ti$. 

\begin{figure}[h]     
%\begin{minipage}{1.08\textwidth}
\begin{center} \vspace{-.35cm}
     \includegraphics[width=7.5cm]{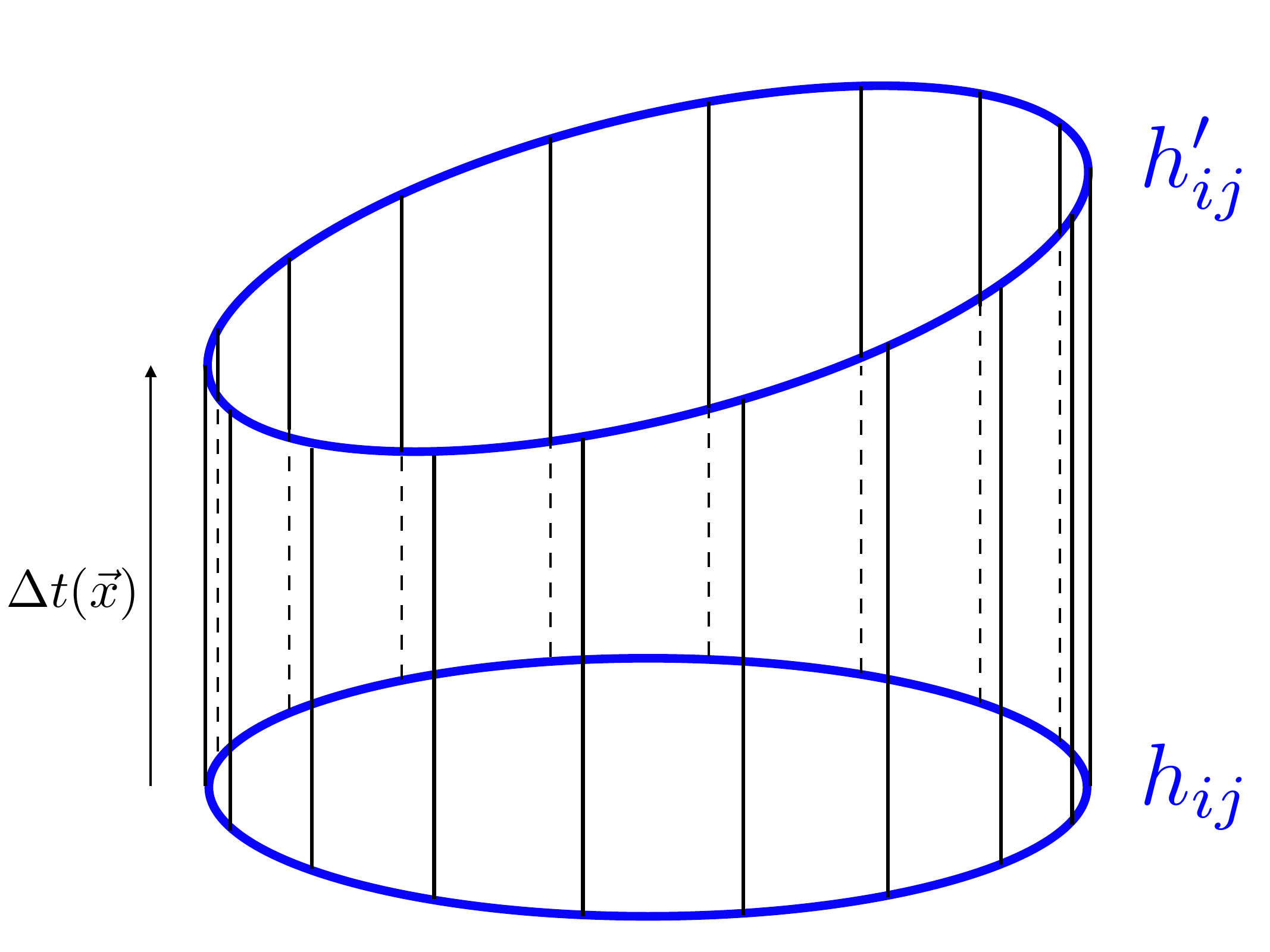}\ \ \ 
        \end{center}\vspace{-.5cm}
  %\end{minipage}
  \captionsetup{width=.95\linewidth}
  \caption{\small The path integral on the RHS of~\eqref{intro_G} extends over all four-geometries that interpolate between the two spatial metrics $h$ and $h'$ (in blue) \emph{in a given time lapse} $\Delta \ti (\vec x)$. The latter can be defined as the proper time at which the geodesic observers that leave $h$ in the orthogonal direction (vertical lines) reach $h'$.} %The proper-time lapse $\Delta t (\vec x)$ is in general dependent on the position of the observers because in the original path integral~\eqref{intro_prima} all possible four-geometries are contained, regardless of the time of arrival. }  
  \label{fig_0}
  \end{figure}

The integral~\eqref{intro_prima} can be done in two steps. First one integrates over all four-geometries that interpolate between $h_{ij}$ and $h'_{ij}$ \emph{in a given time lapse} $\Delta \ti(\vec x)$. This produces an amplitude between $h$ and $h'$  subject to a specific lapse of time $\Delta \ti(\vec x)$,
\begin{equation}
\label{intro_G}
K\bigl(h'_{ij},h_{ij};\Delta \ti\bigr)
= \int \mathcal{D}g \Big|_{\substack{N=1,\,N^i=0, \, \ti_{\rm final}=\Delta \ti(\vec x)}}\ 
e^{\,iI_{\rm EH}[g]}\, .
\end{equation}
In the present construction one can set the initial ADM spatial metric to $\gamma_{ij}(0,\vec x) = h_{ij}(\vec x)$. On each off-shell four-geometry the geodesic observers start their trajectories orthogonal to this hypersurface and  reach their final point on $h'$ in a proper time \(\Delta \ti(\vec x)\). 
The proper-time separation function \(\Delta \ti(\vec x)\) depends in general on \(\vec x\), reflecting the fact that, as boundaries of the four-geometries inside the path integral, $h$ and $h'$ can be tilted with respect to each other in all possible ways (Fig.~\ref{fig_0}).

This means that, in general, the geodesic observers will not be orthogonal also to $h'$ and that the latter cannot be identified with the ADM spatial metric $\gamma_{ij}(t, \vec x)$ at any given time. In the path integral~\eqref{intro_G} $h'$ should thus be imposed as the induced metric at $t = \Delta \ti(\vec x)$.

By summing over all the possible arrival times one recovers the full transition amplitude, 
\begin{equation}
\langle h'_{ij} |  h_{ij}\rangle = \int {\cal D} \Delta \ti \ K(h'_{ij} ,  h_{ij}; \Delta \ti) \, .
\end{equation}

The object defined in~\eqref{intro_G} is analogous to the standard propagator of quantum mechanics. It has the same path integral representation because also in quantum mechanics we restrict the integration to those paths that arrive to destination in a certain time. So, if we restrict to a time interval $\Delta \ti$ which is space-independent,  $K$ is expected to satisfy a standard Schr\"odinger equation,
\begin{equation} \label{intro_prop}
i \partial_\ti K(h'_{ij} ,  h_{ij};  \ti) = {\cal H} \, K(h'_{ij} ,  h_{ij};  \ti)\, ,
\end{equation}
   where ${\cal H}$ is the gravitational Hamiltonian operator written in the variable $h'$. 
   The above handwaving argument leading to eq.~\eqref{intro_prop} has been rigorously developed by Halliwell in the limit of minisuperspace~\cite{Halliwell:1988wc}, the regime considered in this work. As $K$ can be used to propagate a solution of a time dependent wavefunction $\Psi(q^a;t)$, it follows that also the latter  satisfies a standard Schr\"odinger equation, as stated in~\eqref{intro_main}. 
   
   Beyond minisuperspace, a potential issue with the above reasoning is of global type. The \emph{synchronous}  gauge $N=1$ $N^i=0$
 can be associated with a congruence of geodesic observers of fixed spatial coordinates and it breaks down e.g. when two observers of the congruence intersect. 
This can happen in a metric solution of the Einstein equation (i.e. in classical gravity) as well as in the off-shell metrics integrated over in the path integral. In classical as in quantum gravity, (proper) time is a \emph{local} quantity in the sense that it is always well-defined along a worldline while of course it might fail to be defined globally.  
 This point of view is emphasized in the second argument below.

\subsection{The  clock's Hamiltonian point of view} \label{1.2}

One can introduce proper time as a new degree of freedom attached to an observer's worldline $x^\mu(\lambda)$ and governed by a suitable Hamiltonian~\cite{Witten:2023xze} (see also~\cite{Lawrie:2010fg}). Whatever detailed microphysics is used to measure time---typically, some periodic physical process---this should distill into an effective coarse grained degree of freedom $\ti$ whose classical equations of motion reads 
\begin{equation} \label{classicclock}
\frac{d \ti}{d \lambda} = \sqrt{- g_{\mu \nu} \frac{d x^\mu}{d \lambda}\frac{d x^\nu}{d \lambda}}\, .
\end{equation}
In other words, at least classically, $\ti$ should behave as the proper invariant length along the worldline. 
Concretely, this is achieved by attaching to the worldline  a canonical pair $(\ti,p_\ti)$, with action
\begin{equation} \label{actionobs}
I_{\rm clock} = \int_{\gamma} \mathrm{d}\lambda\,
\left[
  p_\ti \,\frac{d\ti}{d \lambda}
  -
  \sqrt{-g_{\lambda\lambda}}\  p_\ti 
\right]\,,
\end{equation}
where $ \sqrt{-g_{\lambda\lambda}}$ is the abbreviation of the expression on the RHS of~\eqref{classicclock}.
Variation of the above with respect to $p_\ti$ clearly enforces~\eqref{classicclock}.

Notice that $\ti$ is really a zero-dimensional quantum mechanical variable because it lives on the observer's worldline. The variance associated to it can be interpreted as the uncertainty about the time shift $\ti \rightarrow \ti  \, + $ constant with which the worldline is embedded in spacetime.\footnote{This point of view should be contrasted with that of a WdW equation without time. Once a field $q^0$ is chosen as ``time" for the standard relational interpretation, one can use a classical solution $q^a(t)$, on which the wavefunction is peaked, to simply translate $q^0$ into $t$~\cite{Piazza:2024lpt}. However, in this case the variable $t$ is merely a re-labeling of $q^0$---it does not introduce any new degree of freedom  nor associated variance.}  Equivalently, it is related to the uncertainty of setting the origin of our clock variable along the worldline. In order to promote $\ti$ to a field one could imagine a dilute gas of observers, each with their own clock, effectively defining a ``time field" everywhere in spacetime. However, in the minisuperspace approach the metric reads, simply, 
\begin{equation}
ds^2 = - N^2 dt^2 + a^2(t) d \vec x^2
\end{equation}
and all fields are reduced to zero-dimensional quantum mechanical variables. In this limit we thus need just one of these clocks and the Hamiltonian that we get from~\eqref{actionobs} becomes 
\begin{equation}\label{intro_clock}
H_{\rm clock} = N p_\ti\,.
\end{equation}

The WdW equation for the combined universe+clock system is simply expressed as (more details on this derivation are given in the appendix) \begin{equation}\label{intro_wdw_clock}
\bigl(\mathcal{H} + p_\ti \bigr)\,\Psi(q^a,\ti)=0\, .
\end{equation}
From this point of view, $\Psi(q^a;\ti)$ is really a standard WdW wavefunction---just that of an extended system that also contains a clock.  Upon use of the canonical relation $p_\ti=-\,i\hbar\,\partial_\ti$, from the above equation one recovers the Schrödinger‑like equation~\eqref{intro_main} in $\ti$. 
\subsection{Energy boundedness and probability density} \label{sec_probability}
One might have noticed that the clock Hamiltonian~\eqref{intro_clock} is unbounded from below. Although the WdW constraint~\eqref{intro_wdw_clock} enforces zero total energy, some solutions could still exhibit arbitrarily large energy exchange between the clock and the universe.  However, in this simplified model, where $H_{\rm clock}$  depends only on $\ti$ and $\cal H$ is $\ti$-independent, no such exchange can occur. More generally, as we are going to discuss, our proposal for time is that it constitutes a component of negligible energy. Preventing a large flow of energy from and to the clock is tantamount to checking the validity of our assumption. 
Another issue is that ${\cal H}$ itself is unbounded, because the conformal mode of the metric (here, the scale factor) comes with the ``wrong sign" of the kinetic term. This is related to the Jeans instability of a gravitational system against collapse. (see also~\cite{Halliwell:1990uy,Giulini:1994dx}). As discussed at the beginning of Sec.~\ref{bounce}, this is the type of instability that can be improved by quantum mechanics. 

The time-dependent wavefunction $\Psi(q^a, \ti)$ then admits the standard interpretation of conditional probabilities \emph{at given time} $\ti$,
\begin{equation} \label{probability}
dP(q^a|\ti) = \sqrt{-g} \, |\Psi(q^a,\ti)|^2 dq^a\, ,
\end{equation} 
where $\sqrt{-g}$ is the invariant volume element in field space (\emph{i.e.} that of the metric $g_{ab}$ in the minisuperspace action~\eqref{action_general} below). $P$ is conserved because it is the charge obtained by integrating the zero-component of the usual conserved current of the Schr\"odinger equation. This straightforward probabilistic interpretation of $\Psi(q^a,\ti)$ is possible because $\Psi$ satisfies~\eqref{intro_main}---an equation different from the WdW equation, which is of  Klein-Gordon type. As such, it avoids the subtleties associated with the conserved WdW probability, such as the appearance of negative probabilities, and will be the interpretation adopted throughout this paper.

In the rest of this paper we apply the Schr\"odinger equation and the probability interpretation above discussed to a variety of situations. 
 Our proposal comes with an additional instruction---that of selecting solutions of~\eqref{intro_main} in which time plays the role of a test field, with negligible backreaction on the metric. We detail this choice in Sec.~\ref{timeastf}. In Sec.~\ref{sec_semiclassical} we show that our formalism produces expected, almost obvious results in the semi-classical regime, where the scale factor is essentially monotonic with respect to time and the wavefunction captures only a piece of evolution of the universe---either contracting or expanding. A smooth ``bounce"  transition between these two phases is recovered if we let the system evolve  through the turnaround. We analyze this bounce numerically in two different scenarios. In the presence of spatial curvature and a cosmological constant we obtain a motion picture of the no-boundary proposal (Sec.~\ref{sec_nb}). For a universe dominated by radiation we also obtain a bounce, that we study in Sec.~\ref{bounce}. We conclude in Sec.~\ref{conclusion} with some speculative outlooks. 

 \section{Time as a test field} \label{timeastf}
 
It looks reasonable to assume that the time field $\ti$ has \emph{negligible back-reaction} on the system-universe. After all, $\ti$ is just a fictitious field that keeps track of some underlying periodic phenomenon.  In the semi-classical limit one writes the total wavefunction as a rapidly oscillating phase, usually controlled by a small parameter $\alpha$ proportional to $\hbar$, times a smoothly modulating piece, whose derivatives are small compared to $\alpha$. In this limit, a \emph{subdominant component} is a field that does not participate to the rapid oscillations.\footnote{Fields with negligible backreaction in the context of the WdW equation have been discussed e.g. in~\cite{Padmanabhan:1988ur,Singh:1989ct}. Vilenkin applies the same approximation to ``small quantum subsystems" in~\cite{Vilenkin:1988yd}. See also the review~\cite{Lehners:2023yrj}. }

To see what this means in practice let us consider a fairly general mini-superspace model
\begin{equation} \label{action_general}
I = \frac{1}{\alpha H_\star} \int d\ti \, \left(\frac{1 }{2} g_{ab} \dot q^a \dot q^b -  H_\star^2 U(q^a)\right). 
\end{equation} 
We are using units $c = \hbar =1$, but we have made explicit some scales that are usually also set to one. Eq.~\eqref{action_general} is obtained by integrating the Einstein Hilbert action over a volume $H_\star^{-3}$ that is supposed homogeneous.  In a decelerating universe we might be interested to follow the backward evolution of a comoving patch of universe entering the Hubble horizon at some epoch. Then we can take $H_\star$ as the Hubble parameter, and $U = {\cal O}(1)$, at that epoch. The dimensionless parameter $\alpha \simeq  G_N H_\star^2 \ll 1$ is the gravitational coupling of this comoving region. 

From~\eqref{action_general} we obtain the Schr\"odinger equation
\begin{equation} \label{wdw}
i \alpha\, \frac{\partial}{H_\star \partial \ti} \Psi(q^a,\ti) =   \left(- \frac{\alpha^2}{2} \mybox  + \,  U (q^a) \right)\Psi(q^a,\ti)\, .
\end{equation} 
It is obvious that $\alpha$ has taken the place of $\hbar$. The box on the RHS is the Laplacian operator of the field-space metric $g_{ab}$.\footnote{Of course, other operator-ordering choices are possible. However we stick with the one because it appears to be to only choice invariant under field redefinition~\cite{Nicolis:2022gzh,Piazza:2024lpt}.}
\vspace{-0.1cm}
\begin{quote}
The {\bf Time as a test field (TTF)} idea suggests to look for solutions of~\eqref{wdw} that, at least in some region of the field space, admit the semiclassical expansion 
\begin{equation}\label{ttf}
\setlength{\fboxsep}{2\fboxsep} \boxed{\Psi(q^a;\ti) \ \sim \ e^{i S(q^a)/\alpha}\ \rchi(q^a)\,  \psi(q^a;\ti)\, .}
\end{equation}
In the above, the functions $S$, $\rchi$ and $\psi$ have analogous variation rates with respect to their arguments, but only $\psi$ depends on $\ti$. They satisfy the following (field space-) covariant equations 
\begin{align} \label{HJ}
& \nabla S \cdot \nabla S + 2 U = 0, \\ \label{chi}
&  \nabla \cdot (\rchi^2 \, \nabla S )= 0, \\ \label{psi}
& \partial_\ti \psi + H_\star \nabla S \cdot \nabla \psi = 0,
\end{align}
up to higher orders in $\alpha$. 
\end{quote}
\vspace{-0.1cm}
The above are obtained by inserting~\eqref{ttf} in~\eqref{wdw}. 
At order ${\cal O}(\alpha^0)$, this yields the Hamilton-Jacobi equation~\eqref{HJ}. The next-to-leading order term, ${\cal O}(\alpha)$, naturally separates into~\eqref{chi} and~\eqref{psi}. This splitting, also discussed by Hartle in Ref.~\cite{Hartle:1992as} at around eq. (9.29), makes the product $e^{i S(q^a)/\alpha} \rchi(q^a)$ an ${\cal O}(\alpha)$  solution of the standard WdW equation (without time), with $\rchi$ the standard WKB pre-factor. Although this choice restricts the potential solutions of eq.~\eqref{wdw}, it appears to preserve the most interesting ones. 

The familiar quantum mechanical setup of eq.~\eqref{wdw} allows us to draw some immediate and  basic properties of the TTF solutions.  
First, the TTF hypothesis~\eqref{ttf} amounts to look for non-stationary states with negligible energy---more precisely, with energy of ${\cal O}(\alpha)$. One way to see this is to consider the  \emph{solution} of the standard Hamilton-Jacobi equation in its separable form,
\begin{equation}
S =  W(q^a) - E\ti\, ,
\end{equation}
with $W$ the \emph{Hamilton characteristic function} and $E$ the energy of the system. When $S$ depends only on the $q^a$ as in~\eqref{ttf}, the energy clearly vanishes in the classical limit, \emph{i.e.} up to order $\alpha$ (or $\hbar$). Moreover,  this condition on the energy applies at all times and beyond the semiclassical regime (e.g. during the bounce of Sec.~\ref{bounce}), simply because, in the case of a time-independent Hamiltonian operator,  the energy is conserved.

\subsection{Comparing \texorpdfstring{$\Psi$}{Psi} and \texorpdfstring{$\Psi_{WdW}$}{Psi_WdW}}

As we are going to see, the semi-classical approximation~\eqref{ttf} typically applies in field-space regions far from the border of field space. This is the case the wavepacket has a the variance of the scale factor $\Delta a$  smaller  than the mean value $\langle a \rangle$. This way, we know that the wavepacket has little support on the border $a = 0$.  In this regime, eq.~\eqref{ttf} can be read as
\begin{equation} \label{vs}
\Psi(q^a;\ti) \ \sim \ \Psi_{WdW}(q^a) \,  \psi(q^a;\ti)\, ,
\end{equation}
in the sense that the combination $e^{iS/\alpha} \rchi$ is a WdW wavefunction to order $\alpha$. So, adding time to the WdW semi-classical picture just means 
adding a gentle modulating factor to the WdW wavefunction, assuring normalizability with respect to the probability interpretation~\eqref{probability}. Among all possibilities, a gaussian ansatz for $\psi$ will be our favorite. It represents a normalizable wavefunction and it is suggested by the negligible non-gaussianity observed in cosmological perturbations.

One point that we try to address in this paper is what happens to the wavepacket once it gets close to the origin at $a=0$. There is clearly a breakdown of the semi-classical regime in this case, but not necessarily of the limit of validity of general relativity if the wavefunctional is still reasonably supported by metrics with sub-Planckian curvatures. In this case, as we are going to show, some type of bounce is generally expected.  After the bounce (or before, if we evolve backward in time) a different semi-classical approximation than~\eqref{vs} applies, which is approximately a time-reversal version of it, i.e. $\Psi_{after}(q^a;\ti) \sim \Psi^*_{WdW}(q^a) \,  \tilde \psi(q^a;-\ti)$. Before looking at the bounce-phase, let us examine the semi-classical regime in a couple of examples. 

\section{The semi-classical regime} \label{sec_semiclassical}

\subsection{Perfect fluid}\label{sec_perfect}

The effective field theory description of a fluid involves three scalar fields and a Lagrangian invariant under internal (to field space) diffeomorphisms~\cite{Soper:1976bb,Dubovsky:2005xd,Endlich:2010hf}. When coupled to gravity, the ``unitary gauge" choice for this system corresponds to using these three fields as the three spatial coordinates. When taking the mini-superspace limit one then ends up with an action for the scale factor only~\cite{Nicolis:2022gzh}, featuring a power-law potential when the equation of state is constant. With an appropriate rescaling of the scale factor $a$ (we are assuming the metric $ds^2 = - d\ti^2 + a^2 d\vec x^2$) the action simply reads
\begin{equation}\label{I2}
I = - \frac{1}{\alpha H_\star} \int d\ti \, \frac12 \left( a \dot a^2  +  H_\star^2\, a^{3 - 2\epsilon}\right)\, .
\end{equation}
Instead of the equation of state parameter $w$ we have used the ``slow-roll" parameter $\epsilon=3(w+1)/2$, which does not need to be small here. During radiation and matter domination, for example, $\epsilon = 2$ and $\epsilon = 3/2$ respectively. 

The classical constraint (Friedmann equation) gives, for an expanding universe,
\begin{equation} \label{constrain}
\dot a = H_\star a^{1-\epsilon} \, .
\end{equation}
As the above equation is of first order, there is no propagating degree of freedom. In this respect, this model shares the same ``rigidity" of de Sitter space, which is obtained from the above in the limit $\epsilon =0$.  The only freedom that we have at the classical level is the choice of initial time when we integrate eq.~\eqref{constrain}.
The situation is paralleled quantum mechanically.
The timeless WdW equation governs some wavefunction $\Psi(a)$ which is difficult to interpret dynamically, as there is no other field than $a$, to use as a reference, or as time. However, adding time at the quantum level necessarily introduces a new degree of freedom. The resulting associated variance can be interpreted as the uncertainty in choosing an initial time for a physical clock. 
 The result is a $\Psi(a,\ti)$ which, at least in the semiclassical approximation, displays the expected classical dynamics as shown in the following. 

In order to apply the semiclassical equations~\eqref{HJ}-\eqref{psi} notice that the field space metric reads $g_{00} = - a$. Equation~\eqref{HJ} gives $S'(a) = \pm \, a^{2 - \epsilon}$. We choose the minus sign, which is appropriate for an expanding universe.\footnote{One can see from~\eqref{I2} that the conjugate momentum to $a$ comes with the ``wrong sign",  $p_a  = - a \dot a /(\alpha H_\star)$. By applying the corresponding operator on the rapidly oscillating phase we obtain 
\begin{equation}
p_a e^{iS/\alpha} = - i \partial_a e^{iS/\alpha} \sim - \frac{a^{2 -\epsilon}}{\alpha} e^{iS/\alpha}\, , 
\end{equation}
consistently with~\eqref{constrain}.}
The equation for $\rchi$ then gives 
\begin{equation}
\rchi = a^{\frac{\epsilon}{2} - \frac34}\, .
\end{equation}

Finally, equation~\eqref{psi} reads 
\begin{equation} \label{eq_psi}
\left(H_\star^{-1} \partial_\ti + a^{1-\epsilon} \partial_a \right) \psi = 0 \, .
\end{equation}
This is solved by any function of the combination $\epsilon H_\star \ti - a^\epsilon $. In particular, we can choose such a function to be a nice Gaussian peaked at zero. This way, $\psi$ is peaked at $a \sim \ti^{1/\epsilon}$, which is the classical behavior. 
\subsubsection*{Radiation}
Let us put all the pieces together by using \emph{radiation domination} ($\epsilon = 2$) as an example. We get to the semi-classical  state
\begin{equation}  \label{wf}
\Psi(a,\ti) \sim  \exp\left[ - \frac{(2 H_\star \ti - a^2)^2}{4 \sigma^2} + \frac14 \ln a - \frac{i a}{\alpha}\right]\, .
\end{equation}
At the exponent  the contributions of $\psi$, $\rchi$ and $S$ appear, in this order. $\sigma$ is the variance of $a^2$. The support of $\Psi$ cannot include the origin $a=0$, so we have to require that $H_\star \ti \gg \sigma$. 

The wavefunction~\eqref{wf} makes a lot of sense. The associated probability~\eqref{probability} is peaked at $a \sim \sqrt{2H_\star \ti}$, the classical behavior of the scale factor, with a small correction coming from the $\ln a$ at the exponent of $\Psi$ and an analogous contribution from the measure $\sqrt{-g} = a^{1/2}$. We can estimate the mean value of $a$ by finding the maximum at the exponent of the resulting probability density,
\begin{equation}
\langle a \rangle = \sqrt{2H_\star \ti} + \frac{\sigma^2}{8 \sqrt{2} \, (H_\star \ti)^{3/2}} + \dots\, .
\end{equation}
Deviations from the classical behavior are proportional to the variance squared $\sigma^2$ and become important at small $\ti$ when, at any rate, we approach the origin $a=0$ and the semiclassical approximation breaks down. We analyze this phase in Sec.~\ref{bounce}. Notice that the model predicts a constant variance for $a^2$, which means that the variance of $a$ itself decreases as $\sim 1/a$. This is also expected. If we consider a classical solution $a\sim \ti^{1/2}$ and we are uncertain about the initial time, this uncertainty propagates in time as $\sim \ti^{-1/2} \sim 1/a$. 
\subsubsection*{Non relativistic matter}
A similar exercise can be done for \emph{matter domination} ($\epsilon = 3/2$). In this case  $S(a) \sim a^{3/2}$, $\rchi = 1$ and~\eqref{eq_psi} is solved by a general function of $3 H_\star \ti  - 2 a^{3/2}$. In summary, the semi-classical wavefunction can be given the gaussian form 
\begin{equation}
\label{matdom}
\Psi(a,\ti) \sim  \exp\left[ - \frac{\left(\frac{3 H_\star \ti}{2} - a^{3/2}\right)^2}{4 \sigma^2}  - \frac{2 i a^{3/2}}{3 \alpha}\right]\, .
\end{equation}
Again, this gaussian ansatz implies that the variance of $a^{3/2}$ is constant in time. 

\subsection{Scalar field and the conservation of $\zeta$}

In order to make contact with cosmological perturbation theory, one might wonder what mini-superspace variable approximates, say, the curvature perturbation on comoving spatial slices  $\zeta$. However, for even defining the $\zeta$ variable, we need to introduce one more degree of freedom into the game.
To this aim, we add a scalar field $\phi$,
\begin{equation}
    I=\frac{1}{\alpha H_{\star}} \int d\ti \, a^3\bigg(-\frac{3\dot{a}^2}{a^2}+\frac{\dot{\phi}^2}{2}-H_\star^2V(\phi)\bigg)\,,\label{eq:action_scalar_field}
\end{equation}
and consider a purely exponential potential $V(\phi)=(3-\epsilon)e^{-\sqrt{2\epsilon}\phi}$.  The field-space metric reads $g_{ab} = \diag(- 6 a, a^3)$. This model admits classical scaling solutions with constant equation of state $w = (2\epsilon -3)/3$,
    \begin{equation}
     \bar a(\ti)=\left(\frac{\ti}{\ti_\star}\right)^{\frac{1}{\epsilon}},\qquad \bar \phi(\ti)=\sqrt{\frac{2}{\epsilon}}\ln\epsilon H_\star \ti. \label{eq:classical_solutions_model1}
\end{equation}
With an appropriate choice of the constant $\ti_{\star}$ we can express the background solution above in a non-parametric ``Hamilton-Jacobi" form, e.g.
\begin{equation}\label{eq:Scaling_Solutionn}
\bar a (\phi) = e^{\phi/\sqrt{2\epsilon}}\, .
\end{equation}

The WdW equation applied to this model can be seen as a quantum version of the Hamilton-Jacobi approach~\eqref{eq:Scaling_Solutionn}. With the WdW wavefunction $\Psi(a,\phi)$ one can study e.g. the average value of the variable $a$ and its uncertainty \emph{at given} $\phi$, and viceversa~\cite{Piazza:2024lpt}. 
The Schr\"odinger approach can be seen instead as the quantum version of~\eqref{eq:classical_solutions_model1}. It produces a different wavefunction, $\Psi(a, \phi, \ti)$, where both $a$ and $\phi$ have independent variances (at given $\ti$). As already emphasized, in this Schr\"odinger approach time really represents an additional degree of freedom in the theory.

 The Schrödinger equation \eqref{wdw} reads
\begin{equation}
  i\alpha \, H^{-1}_{\star} \partial _\ti \Psi=-\frac{\alpha^2}{2}\left[-\frac{1}{6}a^{-2}\partial_a\left(a^{-2}\partial_a\right)+a^{-3}\partial_\phi^2\right]\Psi+(3-\epsilon)a^3e^{-\sqrt{2\epsilon}\phi}\Psi\,.
\end{equation}
It is straightforward to show that under a WKB ansatz of type \eqref{ttf}, 
\begin{equation}
    \Psi(a,\phi,\ti)=\exp\left[\frac{iS(a,\phi)}{\alpha}\right]\rchi(a,\phi)\psi(a,\phi,\ti)\, ,
\end{equation} 
the leading-order HJ equation \eqref{HJ} is satisfied for 
\begin{equation}
    S(\rho,\phi)=-2a^3e^{-\sqrt{\frac{\epsilon}{2}}\phi}.
\end{equation}
This is not the most general solution of the HJ equation, but it is the one naturally associated with the classical trajectory~\eqref{eq:Scaling_Solutionn}, in that the gradient of $S$ is proportional to the flow along the trajectory~\cite{Piazza:2024lpt}. 
The remaining equations \eqref{chi} and \eqref{psi} yield
\begin{align} \label{eq:sch_like_scalar_field}
 a\, \partial_a\rchi+\sqrt{2\epsilon}\, \partial_\phi\rchi+\left(\frac{3-\epsilon}{2}\right)\rchi &= 0\,,\\ \label{eq:sch_like_scalar_field2}
   H_{\star}^{-1}\partial_\ti\psi+e^{-\sqrt{\frac{\epsilon}{2}}\phi}\left(a\, \partial_a\psi+\sqrt{2\epsilon}\, \partial_\phi\psi\right)&=0\,. 
\end{align}  

These equations can be simplified by substituting the field variable $a$ with the combination
\begin{equation}
\zeta = \ln a - \frac{1}{\sqrt{2\epsilon}} \phi\,,\\
 \end{equation}
 while leaving $\phi$ and $\ti$ unchanged. $\zeta$ is a perturbation variable, in the sense that it vanishes on the classical solution~\eqref{eq:Scaling_Solutionn}.
 It is the mini-superspace analogue of the curvature perturbation on comoving spatial slices. 
 Eqs.~\eqref{eq:sch_like_scalar_field} and~\eqref{eq:sch_like_scalar_field2} become independent of $\zeta$,
\begin{align} 
 \sqrt{2\epsilon} \, \partial_{\phi}\rchi + \left(\frac{3-\epsilon}{2}\right)\rchi  &= 0\,,\\
 \partial_{\ti}\psi + H_{\star}\sqrt{2\epsilon}\, e^{-\sqrt{\frac\epsilon2}\phi}  \partial_{\phi}\psi&=0\,. 
\end{align}  
These equations admit solutions of the form
\begin{align}
    \rchi(\phi)&=\exp\left(-\, \frac{3-\epsilon}{2\sqrt{2\epsilon}}\, \phi\right)\,,\\[2mm]
    \psi(\zeta, \phi,\ti)&= f \left(\epsilon H_\star \ti- e^{\sqrt{\frac\epsilon2}\phi}\right) g(\zeta)\, , 
\end{align}
where $f$ and $g$ are generic functions. 

Here again we are happy to recover expected results. By choosing both $f$ and $g$ to be Gaussians centered at the origin, $\psi$ can be made a smooth wavefunction peaked on the classical trajectory~\eqref{eq:classical_solutions_model1}. The time evolution of the variances is particularly simple in the $\zeta$-$\phi$ variables. The variance of $\zeta$ is simply constant, as expected from the classical result that curvature perturbations on comoving spatial slices are conserved on super-Hubble scales (e.g.~\cite{Lyth:2003im,Weinberg:2003sw,Langlois:2005qp}). In the $\phi$ direction, the constant variance is that of $e^{\sqrt{\frac\epsilon2}\phi}$. Such a variance can be seen as an additional quantum uncertainty that has no analogue in standard perturbation theory and that we have introduced by treating time as a field.  On the classical solution, $e^{\sqrt{\frac\epsilon2}\phi}$ scales as $a^{\epsilon}$, so this is the same uncertainty as that of the fluid model (without scalar field) in the previous subsection (see eq.~\ref{eq_psi}).

\section{The no-boundary universe in motion} \label{sec_nb}

We now consider the simplest model for the no boundary proposal, which is a spatially closed minisuperspace model with a cosmological constant~\cite{Hartle:1983ai}. With an appropriate choice of the coefficients the action reads
\begin{equation}
    I= \frac{1}{\alpha H_{\star}} \int d\ti \, \left[-a \dot{a}^2+ H_\star^2 \left(a - a^3 \right)\right] \label{noboundac}\, .
\end{equation}
The solution to the classical constraint equation $\dot a^2 + H_\star^2(1 - a^2)=0$ is (global-) de Sitter space, 
\begin{equation} \label{dsmetric}
ds^2 = H_\star^{-2}\left(-d\ti^2 + a_{ cl}^2 d\Omega\right) \qquad {\rm with} \qquad a_{cl}(\ti) = \cosh (H_\star \ti).
\end{equation}
 Here $H_\star^{-1}$ is the de Sitter radius and, as before, $\alpha \simeq G_N H_\star^2$. As we are now dealing with a closed universe, $a$ really represents the (rescaled) size of an entire spatial slice of this universe, rather than that of a comoving patch. 

At very early times the spatial curvature ($\sim a$) term in~\eqref{noboundac} dominates. When approximating the wavefunction with the saddle point, $\Psi \sim e^{i I[a_{cl}]}$, one can then set the boundary conditions in the Euclidean by demanding that $a_{cl}$ represents a smooth closed geometry~\cite{Hartle:1983ai}. One should then evaluate the on-shell action
\begin{equation}
I^{\rm on-shell} =  \frac{2}{\alpha}\int d\ti' H_\star (a_{cl}^3 - a_{cl}) 
\end{equation}
 along the Euclidean time contour $\ti' = i \pi/2 \rightarrow 0$ followed by the Lorentzian piece $\ti' = 0 \rightarrow \ti$.
The result is the celebrated no boundary wavefunction 
\begin{equation}
\Psi_{NB}(a)  \sim e^{ \frac{2}{ 3 \alpha}[1 - i (a^2 - 1)^{3/2}]}\, .
\end{equation}

%%%%%%%%%%%%%%%%%%%%
\begin{figure}[ht]
%\vspace{-1cm}
\begin{minipage}{1.08\textwidth}
\begin{center} \hspace{-1.9cm}
\includegraphics[width=8cm]{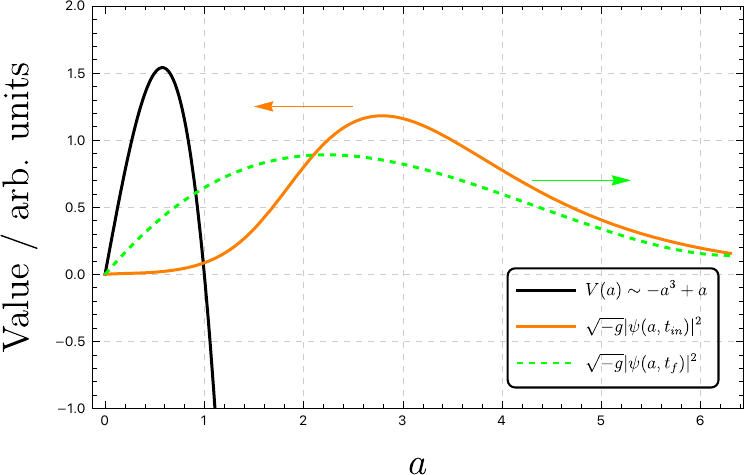}\quad
\includegraphics[width=7.9cm]{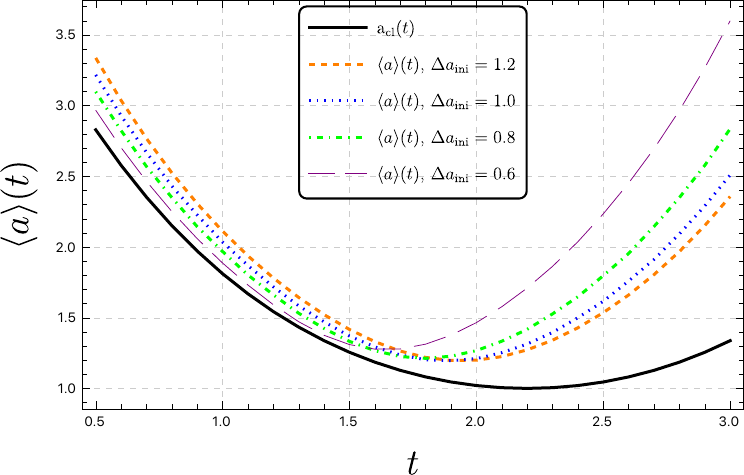} \\[5mm] \hspace{-1.8cm}
\includegraphics[width=7.9cm]{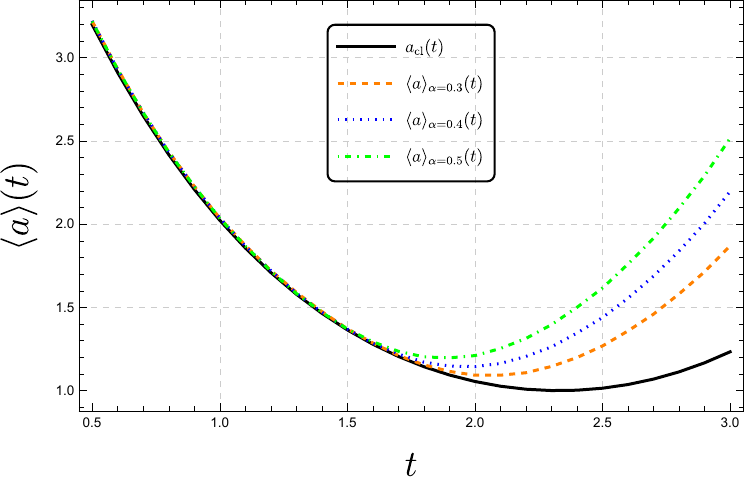}\quad
\includegraphics[width=7.9cm]{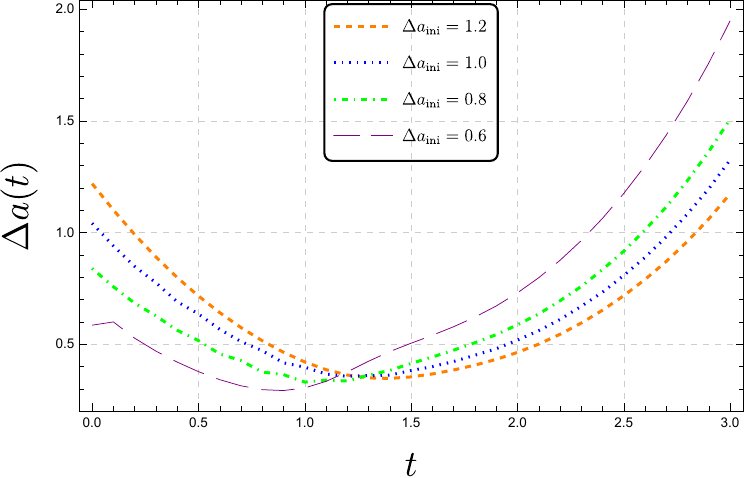} \end{center}%\vspace{-.6cm}
\end{minipage}
\captionsetup{width=1\linewidth}
\caption{\small Top-left panel: representation of the no boundary proposal ``in real time". The state~\eqref{nbpacket} (in orange) is numerically evolved backward in time. After bouncing off the potential barrier (in black) it re-expands (dashed, green curve). Dirichelet boundary conditions are imposed at the origin. The behavior of $\langle a \rangle$ is shown in the bottom left panel (for different values of $\alpha$) and in the top right one (for different values of the initial variance). The classical behavior of $a$ in de Sitter space is given by the thick black curve. The bottom right panel follows instead the evolution in time of the variance of the wavepacket.   } \label{fig_1}
 \end{figure}
 
From the point of view of our Schr\"odinger equation, the above is the leading order WKB approximation $\sim e^{iS/\alpha}$ of the entire wavefunction. It is purely oscillating in the asymptotic $a>1$ region and exponentially suppressed in the classically forbidden region. The corresponding HJ function and WKB prefactor obtained by solving eq.~\eqref{chi} read, respectively 
\begin{equation}
S = - \frac23 (a^2 - 1)^{3/2}, \qquad \rchi =  (a^3 - a)^{-1/4}\, .
\end{equation}
Meanwhile, eq.~\eqref{psi} is satisfied if $\psi(a,\ti)$ is a generic function of the combination $ H_\star \ti -{\rm arcosh}(a)$ (Notice that the field-space metric corresponding to~\eqref{noboundac} is $g_{aa}=-2a$). By putting all together and assuming the usual Gaussian ansatz for $\psi$ we obtain 
\begin{equation} \label{nbpacket}
\Psi \sim \exp\left[ - \frac{\left( H_\star \ti - {\rm arcosh}(a)\right)^2  }{2 \sigma} - \frac14\log (a^3 - a) - \frac{2 i}{3 \alpha}(a^2 - 1)^{3/2} \right] \, .
\end{equation}
The above approximation is valid in the semi-classical limit at $a\gg 1$ and $\ti>0$. We evolve it backward in time numerically by integrating the full Schr\"odinger equation associated with~\eqref{noboundac}. In fact, we know already what happens, at least qualitatively, as this is now standard quantum mechanics. The wavepacket has negligible energy by our TTF assumption, so it simply bounces off the potential back to infinity (Fig.~\ref{fig_1}, top left panel). Notice that in the potential of Fig.~\ref{fig_1}, also a classical particle of zero energy would bounce off to infinity.
This should be contrasted with the radiation bounce of the next section, that does not have a classical analogue.  

Although perhaps unconventional, this bounce interpretation of the no-boundary proposal was clearly emphasized already by Vilenkin in~\cite{Vilenkin:1984wp} (see also~\cite{Danielsson:2021tyb}). It gives a straightforward quantum version of global dS space, with the scale factor shrinking to the dS radius and then expanding again. 
More realistic models of the early universe, with the inflaton already present and the cosmological constant replaced by a slow-roll  potential (e.g.~\cite{Janssen:2020pii,Maldacena:2024uhs}), might not share the same bounce interpretation, once time is properly introduced as a test field. 

\subsection{On quantum de Sitter (dS) space}

Let us briefly comment on the quantum version of dS space that we get in this simple time dependent model. The numerical integration of the full Schr\"odinger equation finds that $\langle a(\ti) \rangle$ departs from the classical solution~\eqref{dsmetric} in that it bounces off more quickly. The classical behavior is recovered  in the limit of infinite variance for the wavepacket, as shown in Fig.~\ref{fig_1}. 

Of course this is just a mini-superspace version of quantum dS space, but there could be some lessons to grab from it. It looks like there is an important difference between static and non-static spacetimes when one is looking for their quantum counterparts. Roughly speaking, we expect the wavefunction of a static spacetime to belong to the same broad class as that of the harmonic oscillator. The latter admits a time independent ground state, which is not less static than a classical particle sitting at the bottom of its potential. The typical example of a static spacetime is \emph{Anti}-de Sitter. By flipping  the sign of the cosmological constant, the potential term  in~\eqref{noboundac} becomes of the confining type. In this case, our (potentially) time-dependent approach has nothing to add to the standard Hartle-Hawking proposal. 

In contrast, time dependent spacetimes, such as dS, seem to belong to a different broad category, that of the \emph{free particle}. Quantum mechanically,  the latter never represents a static situation, because its variance is always evolving in time. In this sense, the state of a free particle is less symmetric than that of an harmonic oscillator, which at least can be static. 
It is not obvious how to even define spacetime symmetries at the level of the gravitational wavefunctional in a background independent-way (see e.g.~\cite{Nicolis:2022gzh}). It is likely, however, that no quantum version of dS can enjoy as many symmetries as its classical counterpart. Even if the average of the scale factor $\langle a \rangle$ behaves as the classical dS solution~\eqref{dsmetric}, its time-evolving variance will break some of the symmetries of  classical dS space. Similar conclusions, from a totally different perspective, were famously drawn in~\cite{Goheer:2002vf}.

\section{The radiation bounce} \label{bounce}

The no-boundary proposal represents a smooth beginning for our universe. 
In our time dependent approach this smoothness is reflected in the way the potential repels the wavepacket from the origin, allowing minimal penetration into the classically forbidden region near the singularity (Fig.~\ref{fig_1}, top left panel).  Interestingly, the same no-boundary potential barrier also prevents a \emph{classical} particle arriving from infinity with negligible total energy from reaching the singularity at $a=0$.  However, stable quantum systems can afford much more singular potentials than that. 

A power-law potential of the type $V\sim - x^{-\beta}$ looks unbounded from below for any $\beta >0$, but the system is perfectly stable quantum mechanically, as long as $\beta <2$. In fact, by applying the Heisenberg principle to the single particle Hamiltonian one gets the lower bound on the energy
\begin{equation}
E = p^2 - x^{-\beta} \ > \ \hbar ^2 x^{-2} - x^{-\beta}\, .
\end{equation}
For $\beta<2$ the above system has a ``Bohr radius", scaling with $\hbar$ as 
\begin{equation} \label{bohrr}
x_{Bohr} \sim \hbar^{\, \frac{2}{2-\beta}}
\end{equation}
and is thus a regular quantum mechanical system.\footnote{In the ``marginal" case $\beta = 2$ the quantum stability of the system depends on the actual numerical coefficient multiplying the potential (see e.g.~\cite{Landau:1991wop}). We refer the reader to~\cite{Piazza:2025uxm} for a throughout discussion on this point.}

The stability of the atom is a genuine quantum effect that has been invoked at times  as an analogy for the   avoidance of the singularities in quantum gravity (e.g.~\cite{Rovelli:2014cta}). Here we  make the quantitative observation that a relativistic fluid corresponds to the case $\beta = 2/3$ and therefore comfortably falls within the class of quantum mechanically stable potentials. In the context of the so-called \emph{unimodular time}, stable quantum bounces have also been studied~\cite{Gielen_2020,Gielen_2022,Gielen_2023,Gielen_2025,Sahota:2025tih}.

\subsection{Radiation domination in flat field-space coordinates}

 In order to see this we use the results of Sec.~\ref{sec_perfect} while referring to the  canonically normalized variable $x \sim a^{3/2}$, as we will do from now on (except in the plots, where we mostly show $a$). In this variable, and up to an irrelevant rescaling, action~\eqref{I2} for $\epsilon = 2$ becomes 
\begin{equation}\label{xvariable}
I = - \frac{1}{\alpha H_\star} \int d\ti \, \frac12 \left(  \dot x^2  +  H_\star^2\, x^{- 2/3}\right)\, .
\end{equation}
The Schr\"odinger equation obtained from this action is that of a particle with the wrong kinetic term in a repulsive potential. However, upon time reversal $\ti \rightarrow -\ti$, this is the equation for a particle with standard kinetic term $p^2 \sim - \partial_x^2$ and attractive potential $V(x) = - x^{-\frac23}$. In order to make connection with the familiar quantum mechanical case, we thus refer from now on to the time-inverted Schr\"odinger equation that we get from~\eqref{xvariable}, 
 \begin{equation} \label{schx}
 i \alpha H_\star^{-1} \partial_\ti \Psi(x, \ti) = - \frac12\left(\alpha^2 \partial_x^2  + x^{-\frac23} \right)  \Psi(x, \ti)\, .
 \end{equation}
 The problem is identical to that of an $S$-wave $\psi_{l=0}(r,\ti) = R(r,\ti)/ r$ scattering off a central potential $\sim - r^{-\frac23}$. In this case,~\eqref{schx} is the equation satisfied by the radial wavefunction $R(r,\ti)$.   The potential is less singular at the origin than that of the hydrogen atom (i.e. $ r^{-\frac23}$ vs. $r^{-1}$), so the standard analysis (e.g.~\cite{Landau:1991wop}) of the wavefunction's boundary conditions applies also here in the very same way.  In particular, the wavefunction near the origin behaves like $\Psi \sim x$  (see also~\eqref{wf2} below). This is the only behavior that guarantees square-integrability and the self-adjointness of the Hamiltonian operator, and thus a unitary evolution for the system.\footnote{By inserting the ansatz $\Psi = x^{\gamma}$ into eq.~\eqref{schx} one sees that the most singular term of the equation goes as $\gamma(\gamma - 1) x^{\gamma-2}$. So, strictly speaking, both $\gamma =1$ \emph{and} $\gamma = 0$ are possible behaviors close to the singularity.  Adopting the Dirichlet boundary condition ($\gamma=1$) is consistent with the De Witt condition~\cite{DeWitt:1967yk}, which reflects the philosophy that the wave function must vanish at the classical singularity to ensure the probability of finding the universe in a singular state is zero. While $\Psi \sim$ const. boundary conditions could be explored in this case, they should generally be rejected because they belong to the ``wrong branch" of asymptotic behaviors, in the sense that those solutions become singular for generic angular momentum $l \neq 0$ (in the case of a particle in a central potential)~\cite{Landau:1991wop} and for generic number of dimensions/fields~\cite{Piazza:2025uxm}.   } 
 
While  imposing Dirichlet boundary conditions to avoid the singularity might appear circular, this method would fail if not for the relatively mild singular behavior of the potential at the origin, as explained at the beginning of this section. Such a behavior is characteristic of radiation interacting with gravity. Since "radiation" can be understood as the dominant state of matter at very high densities/curvatures, the stability of this potential could represent an intriguing self-correcting mechanism of the system gravity+matter  to prevent the singularity.  
Of course, this is just wishful thinking until we are not able to assess the UV sensitivity of this mechanism to the physics at the Planck scale.   We will turn to this point, albeit not conclusively, at the end of this section. 
 
 \begin{figure}[ht]
\begin{minipage}{1.08\textwidth}
\begin{center} \hspace{-1.9cm}
\includegraphics[width=8cm]{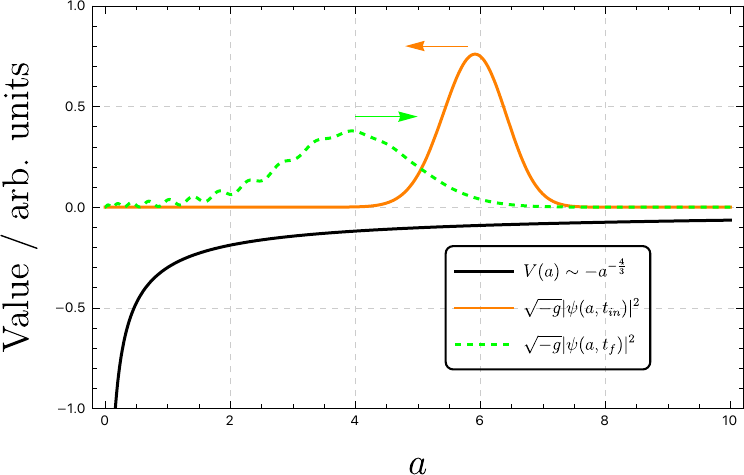}\quad 
\includegraphics[width=7.9cm]{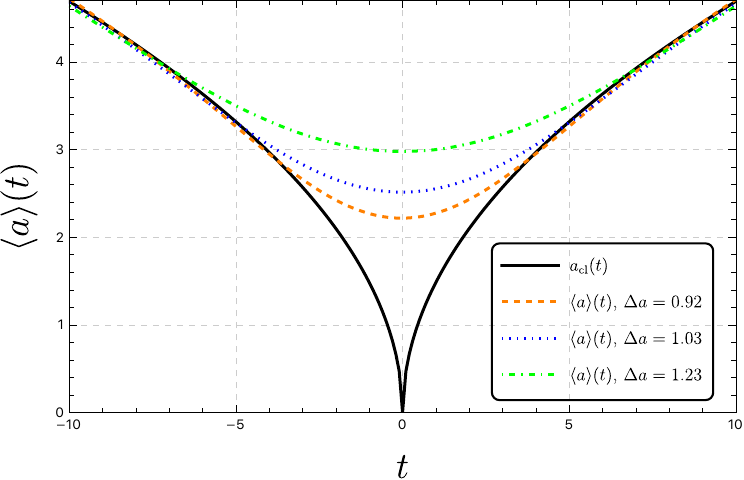} \\[5mm] \hspace{-1.9cm}
\includegraphics[width=8cm]{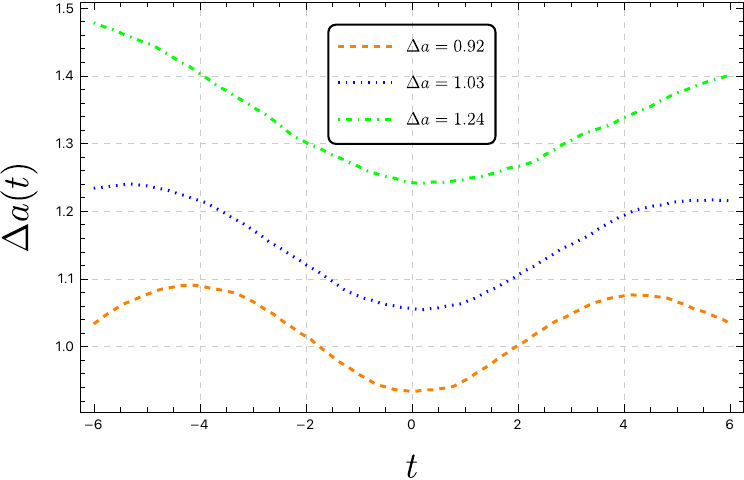}\quad 
\includegraphics[width=7.9cm]{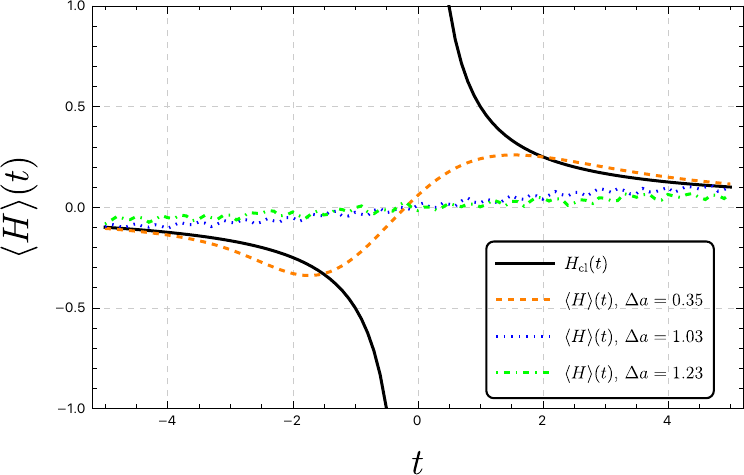} \quad 
\hspace{-1.8cm}
\end{center}%\vspace{-.5cm}
\end{minipage}
\captionsetup{width=.95\linewidth}
\caption{\small Top-left panel: pictorial representation of the radiation bounce. The behavior of $\langle a \rangle$ is shown in the top right panel for different initial values of $\Delta a$. The evolution of the spread of these solutions is shown in the bottom-left. Bottom-right: the behavior of $\langle H \rangle$  can be made arbitrarily smooth by choosing a large initial variance. The red curve has been chosen to represent a ``more singular" solution than in the other panels for pictorial clarity.   }  \label{fig_2}
\end{figure}
 
\subsection{Features of the bounce}
We set initial conditions for the bounce in the semiclassical regime (eq.~\ref{wf}),
\begin{equation}  \label{wf2}
\Psi(x,\ti) \sim  \exp\left[ - \frac{\left(H_\star \ti - \frac34 x^{\frac43}\right)^2}{4 \sigma^2} + \frac16 \ln x - \frac{3i }{4\alpha}x^{\frac23}\right]\, .
\end{equation}
While the above is accurate at $H_\star \ti  \gg \sigma$, we numerically solve~\eqref{schx} backward in time to enquire about the behavior close to the origin. 
To get more analytical insight we can also expand~\eqref{schx} at the origin. One finds that, compatibly with the Dirichlet boundary conditions, the solution admits the double expansion
\begin{equation} \label{expansion}
\Psi(x,\ti) = \frac{616}{81} \alpha^4  f(\ti) \, x -  \frac{44}{9} \alpha^2 f(\ti) \, x^\frac73 - i  \frac{616}{243} \alpha^3 \dot f(\ti) \, x^3 + f(\ti) \, x^\frac{11}{3} + \, \dots
\end{equation}

Figure~\ref{fig_2}, top-right panel, shows the behavior of $\langle a \rangle$ during the bounce. While initially following the contracting classical trajectory $a_{cl}\sim (-\ti)^{1/2}$,  $\langle a \rangle$ detaches from it and reaches a minimal value $\langle a \rangle_{min}$ before re-expanding. Either of two mechanisms are at play that determine the value of $\langle a \rangle_{min}$.
One can try to optimize the variance of the wavepacket just before the bounce to make  $\langle a \rangle_{min}$ as small as possible. In this regime, we find empirically that $\langle a \rangle_{min}$ is a certain function of  $\alpha$.\footnote{Contrary to initial expectations, $\langle a \rangle_{min}$  does not scale as the Bohr radius~\eqref{bohrr}, i.e. as $x_{Bohr} \sim \alpha^{3/2}$, $a_{Bohr} \sim \alpha$. The point is that, under this naive scaling, the energy would scale as $E_{Bohr}\sim \alpha^{-1}$. This contradicts    the TTF hypothesis which implies that  $\Psi(a,\ti)$ should have energy of ${\cal O}(\alpha)$. }  If instead the variance $\Delta a$ is set sufficiently large, the value of $\alpha$ becomes irrelevant and  $\langle a \rangle_{min}\sim \Delta a$ (Table~\ref{tab:side-by-side}).  
By choosing the incoming variance arbitrarily large, the bounce can be made arbitrarily smooth (Fig.~\ref{fig_2}, right panels). 

By smooth here we mean for example a small value of the time derivative of $\log \langle a\rangle$. 
One can even attempt to define a \emph{Hubble parameter operator} $H$. Using the fact that, classically, $ \dot x = \alpha H_\star \, p_x $ one can guess the following expression for $H$ (up to an order-one constant), 
\begin{equation}
H = -\frac{i \alpha H_\star}{2}\left(x^{-1} \partial_x + \partial_x x^{-1}\right)\, .
\end{equation}
This operator is at least symmetric on the domain ${\cal L}^2(0,+\infty)$ with Dirichelet boundary conditions at the origin. Despite it is not self-adjoint, the formal expectation value of $H$ on our wavefunctions is finite near the origin in virtue of~\eqref{expansion}. We have numerically evaluated it and plotted in Fig.~\ref{fig_2} (bottom-right panel).

\begin{table}[htb]
\centering
% Left table
\begin{minipage}{0.48\linewidth}
\centering
\caption*{$\langle a \rangle_{min}$}%\vspace{-2mm}
\setlength{\tabcolsep}{8pt}
\begin{tabular}{|l|l|l|l|}
\hline
\diagbox{$\alpha$}{$\Delta a_{\text{ini}}$}
& 0.89 & 1.02 & 1.31 \\
\cline{1-3}
\arrayrulecolor{black}\cline{4-4} % Red line
1   & 2.35 & 2.65 & 2.95 \\
\arrayrulecolor{black}\cline{3-3} % Red line above column 2
0.5 & 2.25 & 2.50 & 2.95 \\
\arrayrulecolor{black}\cline{2-2} % Red line above column 2
0.1 & 2.20 & 2.50 & 2.95 \\

0.05& 2.20 & 2.50 & 2.95 \\

0.01& 2.20 & 2.50 & 2.95 \\

\arrayrulecolor{black}\hline
\end{tabular}
\end{minipage}\hfill
% Right table
\begin{minipage}{0.48\linewidth}
\centering
\caption*{$\sim \Delta a_{bounce}$} \vspace{-2mm}
\setlength{\tabcolsep}{8pt}
\begin{tabular}{|l|l|l|l|}
\hline
\diagbox{$\alpha$}{$\Delta a_{\text{ini}}$}
& 0.89 & 1.02 & 1.31 \\
\hline
1   & 1.20 & 1.30 & 1.50 \\

0.5 & 1.01 & 1.16 & 1.30 \\
0.1 & 0.92 & 1.03 & 1.23 \\

0.05& 0.92 & 1.03 & 1.22 \\

0.01& 0.92 & 1.03 & 1.22 \\
\arrayrulecolor{black}\hline
\end{tabular}
\end{minipage}
  \captionsetup{width=.95\linewidth}
\caption{\small The minimal value reached by $\langle a\rangle$ (left) and the value of $\Delta a$ close to the bounce (right) for different choices  of $\Delta a_{\text{ini}}$ and $\alpha$.  
Below the horizontal lines $\langle a\rangle_{min}$ becomes insensitive to $\alpha$. }
\label{tab:side-by-side}
\end{table}

 \subsection{The limitations of mini-superspace, anisotropy and UV sensitivity}

The radiation bounce described above is encouraging in several respects and may have implications for spacetime singularities and cosmology. For these applications, however, the present model also exhibits clear caveats and limitations.

While radiation is already the most singular type of ``matter" and should thus prevail close to the collapse,  
the very mini-superspace approximation that we are adopting is extremely restrictive (see e.g.~\cite{Nicolis:2022gzh}) and should eventually be improved into a more realistic field-theory analysis, capable of encoding the effects of  inhomogeneities  and anisotropies.  A glimpse of optimism, in this respect, comes from the fact that, at least classically, anisotropy is the component that tends to dominate when approaching the singularity. 
By generalizing the classical analysis of the seminal work of Belinsky, Khalatnikov  and Lifshitz~\cite{Belinsky:1970ew}, one may attempt to approach the singularity along the worldline of a freely falling observer and argue that, as in the classical case, the spacetime in the vicinity of such an observer is well approximated by a homogeneous---though generally anisotropic---universe (this hypothesis has been largely confirmed by numerical analysis, e.g.~\cite{Garfinkle:2003bb} and referenced therein). This suggests in order to investigate the singularity, one may first incorporate just anisotropies while leaving inhomogeneities aside. This represents a natural and relatively straightforward extension of the present model, an analysis that is carried out by one of us in a companion paper~\cite{Piazza:2025uxm}.

On another note, our model is completely oblivious to UV physics, as we are using a straightforward quantum version of general relativity, a natural question is thus---how sensitive is this mechanism to the UV physics beyond GR? 

The scenario that we find most compelling is when the wavepacket approaches the bounce with a large variance $\Delta a$. In this regime, as we have discussed, the system seems insensitive to the value of $\alpha$ and $  \langle a \rangle_{min}\!\simeq \Delta a$. This is the expected behavior of a particle which cannot localize closer to the origin than the extent of its spread. 
On the other hand, the system cannot be totally insensitive to $\alpha$,  as we know that the corresponding classical system is unstable. The stability of our bounce is still a quantum ($\hbar \neq 0$) effect, so the unknown UV physics should be at least kind enough not to spoil such a stability. Once this is achieved,  one might hope that the specific details of the UV completion are ultimately irrelevant, and the main features of this scenario are essentially captured already by the low-energy theory. 

Also, one should distinguish the possibly highly oscillatory behavior of the wavefunction close to bounce from that of the metrics on which the wavefunction has support. 
In some more realistic models beyond the mini-superspace approximation, it would be interesting to investigate the structure of the gravitational wavefunctional during the bounce. The main average quantities, $\langle a \rangle$ and $\langle H \rangle$, look extremely smooth in our simple model. However, it would be important to understand how much support the wavefunctional has on metrics that have trans-Planckian curvatures. One should hope this support to be negligible, so that we are still in the domain of validity of general relativity, despite encountering phenomena that are not accounted for by the classical equations of motion.

On another note, our model is completely oblivious to UV physics, as we are using a straightforward quantum version of general relativity, a natural question is thus---how sensitive is this mechanism to the UV physics beyond GR? 

The scenario that we find most compelling is when the wavepacket approaches the bounce with a large variance $\Delta a$. In this regime, as we have discussed, the system seems insensitive to the value of $\alpha$ and $  \langle a \rangle_{min}\!\simeq \Delta a$. This is the expected behavior of a particle which cannot localize closer to the origin than the extent of its spread. 
On the other hand, the system cannot be totally insensitive to $\alpha$,  as we know that the corresponding classical system is unstable. The stability of our bounce is still a quantum ($\hbar \neq 0$) effect, so the unknown UV physics should be at least kind enough not to spoil such a stability. Once this is achieved,  one might hope that the specific details of the UV completion are ultimately irrelevant, and the main features of this scenario are essentially captured already by the low-energy theory. 

Also, one should distinguish the possibly highly oscillatory behavior of the wavefunction close to bounce from that of the metrics on which the wavefunction has support. 
In some more realistic models beyond the mini-superspace approximation, it would be interesting to investigate the structure of the gravitational wavefunctional during the bounce. The main average quantities, $\langle a \rangle$ and $\langle H \rangle$, look extremely smooth in our simple model. However, it would be important to understand how much support the wavefunctional has on metrics that have trans-Planckian curvatures. One should hope this support to be negligible, so that we are still in the domain of validity of general relativity, despite encountering phenomena that are not accounted for by the classical equations of motion. 

\section{Comments and speculative outlooks} \label{conclusion}

Let us take the radiation bounce seriously for a moment as a possible starting point for our universe. One may want to find a way to attach a subsequent inflationary phase to it, to ensure the observed homogeneity, isotropy, and the primordial power spectrum of cosmological perturbations. Then reheating would be needed to enter the standard radiation era. 
However, before diving into model-building,  it would be interesting to understand whether the radiation bounce alone already provides some of the key benefits of inflation with the minimal possible ingredients (see also~\cite{Boyle:2018tzc} for reasoning along these lines).  During the bounce itself, $\langle \ddot a \rangle >0$, so this is technically an inflationary phase, although relatively  short-lived. More generally, as far as homogeneity and isotropy are concerned, cosmic bounces are known to potentially establish the ``right kinematics" for perturbation modes, allowing them to exit the horizon before the bounce and re-enter during the standard decelerating phase (e.g.~\cite{Gasperini:2002bn,Lehners:2008vx,Khoury:2008wj}). 

However, the radiation bounce that we have discussed also represents an important deviation from the standard geometrical picture of GR, so some of these classical characterizations (e.g. ``enter and exit the horizon") should be revisited. 
As we have seen our bounce necessarily involves a \emph{highly quantum phase}, with the variance of the scale factor comparable to the scale factor itself.  At the same time, the mean of the Hubble parameter remains well below the Planck scale, which suggests that the wavefunction is still mostly supported on smooth metrics.  Such \emph{low-energy/large-variance} situations, are very fascinating because they hint to macroscopic quantum gravitational effects which are under the control of the low-energy theory.  These situations---still to be fully understood---are commonly associated with  a breakdown of the standard geometric description, but they do not directly imply any localized ``firewall" or singularity. As pointed out e.g. in~\cite{Piazza:2021ojr,Piazza:2022amf,Nitti:2024iyj} deviations from standard geometry and standard causality happen at large separation.  One  diagnostic of such highly quantum phases is represented by \emph{average distances}~\cite{Piazza:2021ojr}, which keep smooth but fail to satisfy a standard property (additivity) of geodesic distances. Clearly, in such a highly fluctuating metric, causality relations should be reconsidered~\cite{Piazza:2022amf}, and it is not excluded that this quantum phase could act as a powerful \emph{scrambling} and \emph{homogenizing} mechanism for the relevant degrees of freedom. A better comprehension of these aspects and their implications on cosmology is clearly needed and is left for future work.

\section*{Acknowledgments} We acknowledge conversations and exchanges with  Eugenio Bianchi, Pietro Don\`a, Angelo Esposito, Ben Freivogel,  Maurizio Gasperini, Steffen Gielen, Jean-Luc Lehners,   Alberto Nicolis, Patrick Peter, Alessandro Podo,  Riccardo Rattazzi, Sergey Sibiryakov, Alexander Taskov, Gabriele Veneziano and especially Andrew Tolley. This work received support from the French government under the France 2030 investment plan, as part of the Initiative d'Excellence d'Aix-Marseille Universit\'e - A*MIDEX (AMX-19-IET-012). It was also supported by the ``action th\'ematique" Cosmology-Galaxies (ATCG) of the CNRS/INSU PN Astro and by  the Agence Nationale de la Recherche under the grant ANR-24-CE31-6963-01.

\appendix

%\section*{Appendix}
%More formally, before taking the mini-superspace limit, one can start with the total action of the system gravity + matter + observer + clock which, in the Hamiltonian ADM formalism, reads
%{\small
%\[ I = \int dt d^3x \left( \pi^{ij} \dot{\gamma}_{ij} + p_{\phi}\dot{\phi} - N (\mathcal{H}_{G} + \mathcal{H}_{mat}) - N^i (\mathcal{H}^G_i + \mathcal{H}^{mat}_i) \right) + \int_{\gamma} d\lambda \left( p_\ti \frac{d\ti}{d\lambda} - \sqrt{-g_{\lambda\lambda}} ( p_\ti + m) \right). \]}
%In this setup the trajectory of the observer, $x^\mu(\lambda)$ is also a dynamical variable, which up to the small clock corrections follows a geodesic, due to the mass term $m$. For simplicity we have consider a standard scalar field as matter. When we truncate to mini-superspace, homogeneity sets the shift vector to $N^i = 0$, the lapse to $N = N(t)$, the spatial metric $h_{ij} = a^2(t) \delta_{ij}$ and $\phi(x^\mu) = \phi(t)$. The observer's trajectory can be set to be $x^\mu = \delta^\mu_0 t$ and it is not a dynamical variable any longer, so the mass term can be omitted. At the same time,  $\sqrt{-g_{\lambda\lambda}}\rightarrow N \frac{dt}{d\lambda}$. Variation with respect to $N$ then yields the integrated total constraint $N(\mathcal{H} + p_\ti) = 0$.

\section{Details on the mini-superspace limit of gravity +  clock}

The total action of the system gravity + matter + observer + clock, in the Hamiltonian ADM formalism, and in the full theory, reads {\small \[ I = \int dt d^3x \left( \pi^{ij} \dot{\gamma}_{ij} + p_{\phi}\dot{\phi} - N (\mathcal{H}_{G} + \mathcal{H}_{mat}) - N^i (\mathcal{H}^G_i + \mathcal{H}^{mat}_i) \right) + \int d\lambda \left( p_\ti \frac{d\ti}{d\lambda} - \sqrt{-g_{\lambda\lambda}} ( p_\ti + m) \right), \]} where the second integral is taken along the trajectory  of the observer, $x^\mu(\lambda)$. In this setup, the latter is  also a dynamical variable. Classically, it satisfies the geodesic equation in virtue of the mass term $m$ and up to the (assumed, small) corrections due to the coupling with the clock's degrees of freedom. For simplicity we consider here a standard scalar field as matter, but the same logic applies for other fluid-dominant components. 

When we truncate to minisuperspace, we restrict the configuration space to metrics and fields that are spatially homogeneous. This "freezes" all but a finite number of degrees of freedom, the shift vector is set to $N^i = 0$, the lapse to $N = N(t)$, the spatial metric to $h_{ij} = a^2(t) \delta_{ij}$, and the matter to $\phi(x^\mu) = \phi(t)$. The fields are eventually reduced to zero-dimensional variables by integrating the spatial part of the action over a comoving volume $V = H_{\star}^{-3}$. This spatial integration step is what defines the dimensionless gravitational coupling $\alpha \simeq G_N H_{\star}^2$.

At the same time, the observer's trajectory can be set to $x^\mu = \delta^\mu_0 t$ and is no longer a dynamical variable, allowing us to omit the mass term $m$. Under these conditions, the proper time rate simplifies to $\sqrt{-g_{\lambda\lambda}} = N \frac{dt}{d\lambda}$, and the entire action can be written as an integral over $t$,
\[ I \sim \int dt \left( p_a \dot{a} + p_\phi \dot{\phi} + p_\ti \dot{\ti} - N (\mathcal{H}_G+\mathcal{H} _{mat}+ p_\ti) \right). \]
Varying this action with respect to the lapse $N$ enforces the integrated total constraint $(\mathcal{H} + p_\ti) = 0$. Upon quantization, identifying the clock momentum with the operator $p_\ti = -i\hbar\partial_\ti$ recovers the Schrödinger-like evolution of~\eqref{intro_main} .

\renewcommand{\baselinestretch}{1}\small
\bibliographystyle{ourbst} %REMOVED DUE TO BIBLATEX
%\bibliography{replicaBib}
\bibliography{references} %REMOVED DUE TO BIBLATEX

\end{document}